# Bank Performance Determinants: State of the Art and Future Research Avenues

**Anas Azzabi[*], Younes Lahrichi**

Department of finance (LAREF), The Higher Institute of Commerce and Business Administration (Groupe ISCAE), Casablanca, Morocco

**ABSTRACT**

*This paper aims to bring an up-to-date and organized review of literature on the determinants of banks performance and suggesting new research avenues. This paper discusses the main approaches that molded the debate on banks' performance and their determinants. Understanding them allows bank managers and regulators to improve the sectors' efficiency and to deal with the new trends of their industry, and academicians to enrich research and knowledge on this field. We show that despite the importance of the existent literature, this paper's subject is still a ripe field for research, especially when considering the evolution of FinTechs and digitalization.*

## Introduction

Bank performance is the capacity of a bank to achieve its objectives, create value for its stakeholders and outperform its competitors (Chenini & Jarboui, 2018). It is affected by specific determinants, such as market concentration, economic growth, regulations etc.

This paper aims to provide the state of the art of the scientific research on bank performance determinants. Secondly, it aims to highlight the gaps observed within the literature, and how future research avenues can address them.

*[*]Correspond author E-mail address: aazzabi_doct26@groupeiscae.ma*

The remainder of this paper is structured as follows: section 1 shows the publishing activity by year. Section 2 shows the top 20 most cited articles in the field of bank performance. Section 3 shows the articles per studied region or country. Section 4 shows the econometric models used by the most cited articles in the literature. Section 5 shows the type of papers based on the empirical approach. Section 6 shows the review of the literature. Section 7 shows the variables used throughout the sampled literature and their proxies. Section 8 shows the future research avenues. Section 9 concludes.

**Publishing activity by year**

Graphic 1 below shows the publication trend in the field of bank performance, according to Alam et al. (2021).

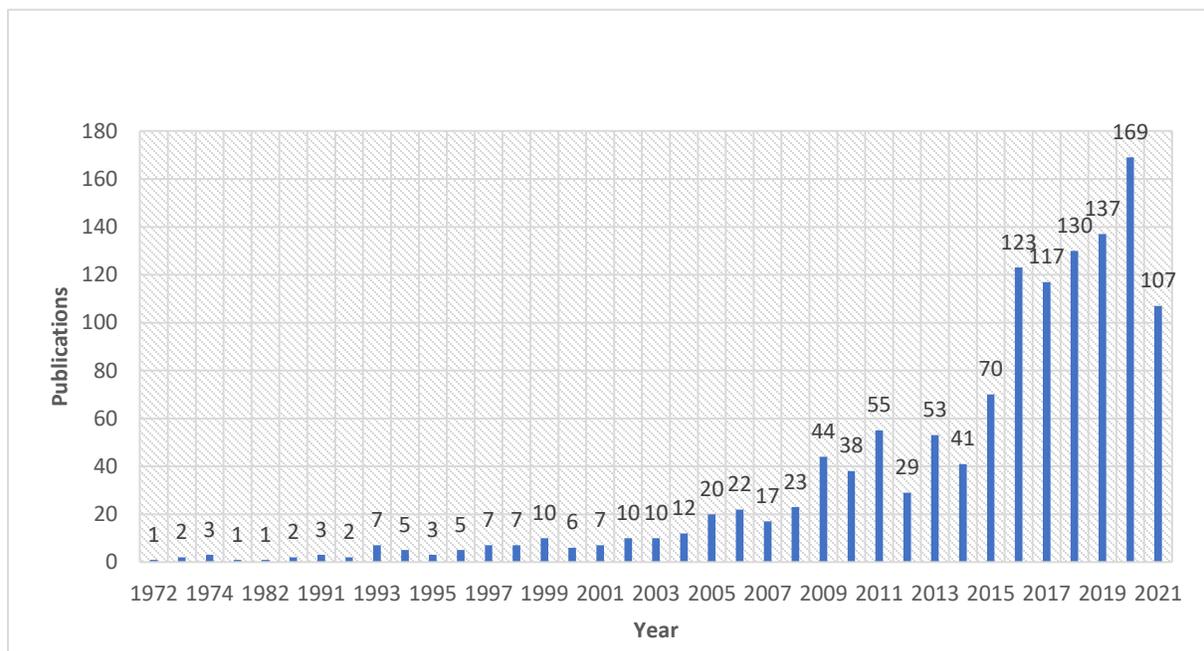

Source: Alam et al. (2021), Biblioshiny R package, Web of Science

*Figure 1.* Publication trend in the field of bank performance

We can conclude that research in the field of bank performance has been documented since the 1970s, but has gained significant interest from the scientific community as of 2009.

**Top 20 most cited articles**

Alam et al. (2021) have compiled a list of the most cited articles on the field of bank performance:

Table 1.
*Top 20 most cited articles*

| Research papers | Authors | Publisher | Citations |
|---|---|---|---|
| Inside the black box: What explains differences in the efficiencies of financial institutions? (1997) | Berger A. N., Mester L. J. | Journal of Banking & Finance | 665 |
| Profitability and marketability of the top 55 US commercial banks (1999) | Seiford L. M., Zhu J. | Management Science | 549 |
| Islamic vs. conventional banking: Business model, efficiency and stability (2013) | Beck T., Demirguc-Kunt A., Merrouche O. | Journal of Banking & Finance | 512 |
| The credit crisis around the globe: Why did some banks perform better? (2012) | Beltratti A., Stultz R. | Journal of Financial Economics | 484 |
| How does capital affect bank performance during financial crises? (2013) | Berger A. N., Bouwman, C. H. S. | Journal of Financial Economics | 463 |
| Bank performance, efficiency and ownership in transition countries (2005) | Bonin J. P., Hasan I., Wachtel P. | Journal of Banking & Finance | 451 |
| Effect of Financial Development on the Transmission of Monetary Policy (2009) | Berger A. N. | Journal of Banking & Finance | 438 |
| Bank CEO incentives and the credit crisis (2011) | Fahlenbrach R., Stulz R. M. | Journal of Financial Economics | 390 |
| Corporate governance in banking: The role of the board of directors (2008) | De Andres P., Vallelado E. | Journal of Banking & Finance | 359 |
| Risk management, corporate governance, and bank performance in the financial crisis (2012) | Aebi V., Sabato G., Schmid M. | Journal of Banking & Finance | 303 |
| A study of bank efficiency taking into account risk-preferences (1996) | Mester L. J. | Journal of Banking & Finance | 286 |
| On the implications of market power in banking: Evidence from developing countries (2010) | Ariss R. T. | Journal of Banking & Finance | 274 |
| Big Bad Banks? The Winners and Losers from Bank Deregulation in the United States (2010) | Beck T., Levine R. | Journal of Finance | 272 |
| Concentration and foreign penetration in Latin American banking sectors: Impact on competition and risk (2007) | Micco A., Yeyati E. L. | Journal of Banking & Finance | 268 |
| Outreach and Efficiency of Microfinance Institutions (2011) | Hermes N., Lensink R., Meesters A. | World Development | 263 |
| The efficiency of financial institutions: A review and preview of research past, present and future (1993) | Berger A. N., Hunter W. C., Timme S. G. | Journal of Banking & Finance | 253 |
| Efficiency and risk in European banking (2011) | Fiordelisi F., Marques-Ibanez D., Molyneux P. | Journal of Banking & Finance | 252 |
| A comparative study of efficiency in European banking (2003) | Casu B., Molyneux P. | Applied Economics | 236 |
| What explains the low profitability of Chinese banks? (2009) | García-Herroro A., Gavilá S., Santbárbara D. | Journal of Banking & Finance | 228 |
| Bank Ownership Reform and Bank Performance in China (2009) | Lin X., Zhang Y. | Journal of Banking & Finance | 225 |

Source: Alam et al. (2021), Biblioshiny R package, Web of Science.

**Research by studied country or region**

Graphic 2 below shows the focus of the sampled literature by studied country or region.

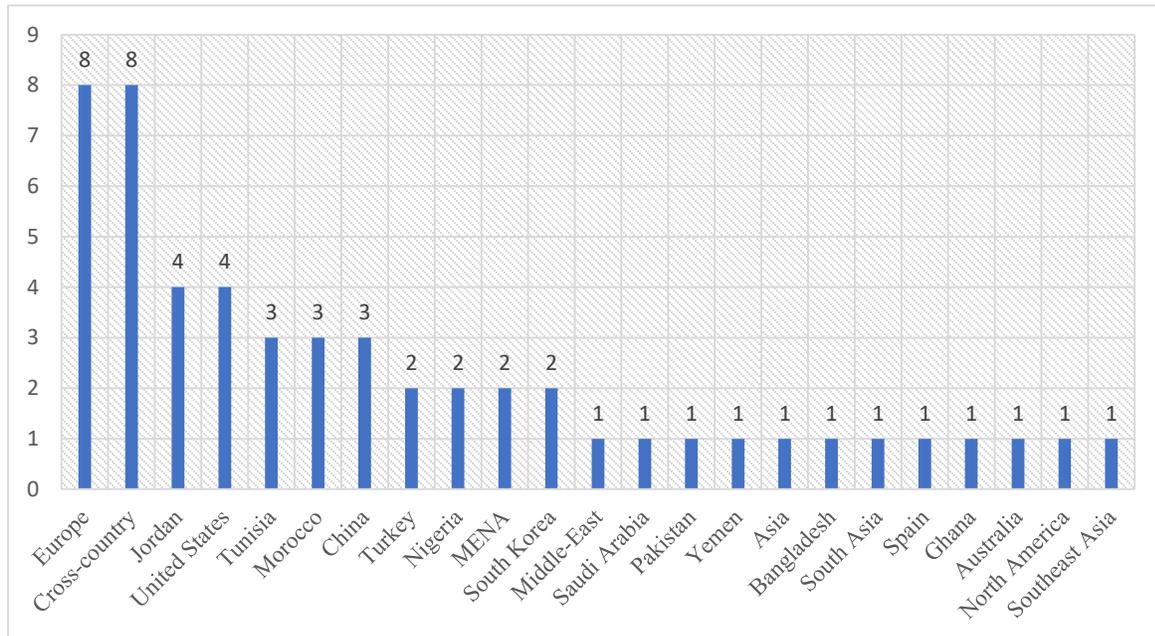

Source: Authors

*Figure 2.* Sampled literature by studied regions/countries

Additionally, graphic 3 shows the studied countries and regions of the 20 most cited articles.

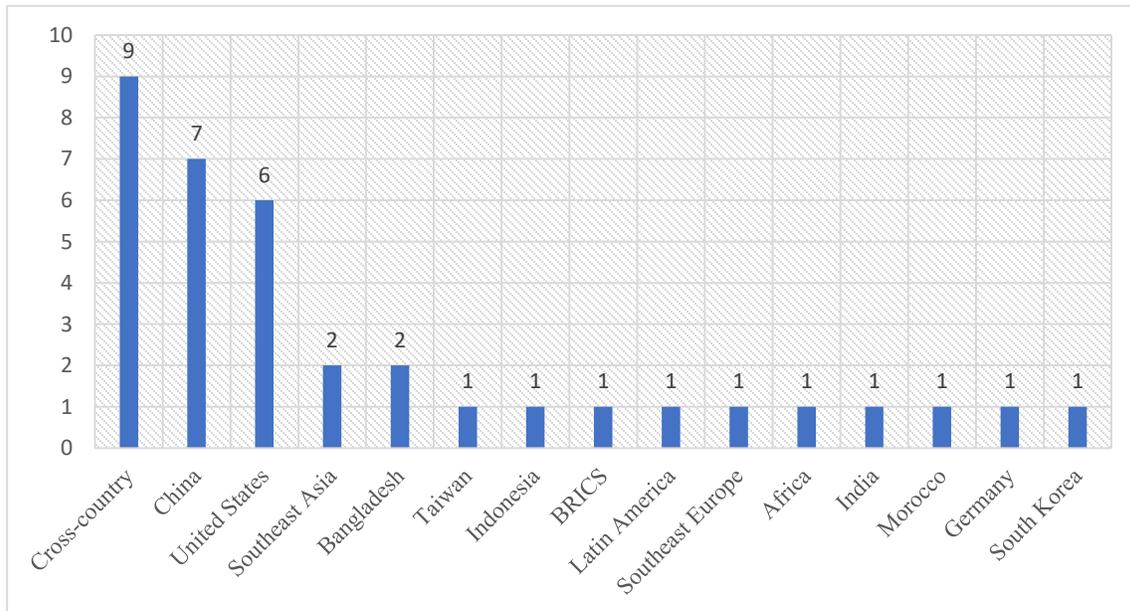

Source: Authors. Data from Alam et al. 2021.

*Figure 3.* 20 most cited articles by studied regions/countries

Different countries were studied by the literature. However, the retained data covers years 1998 through 2018.

**Most used econometric models**

Graphics 4 and 5 below show the econometric model usage by the sampled literature and the top 20 most cited articles, respectively.

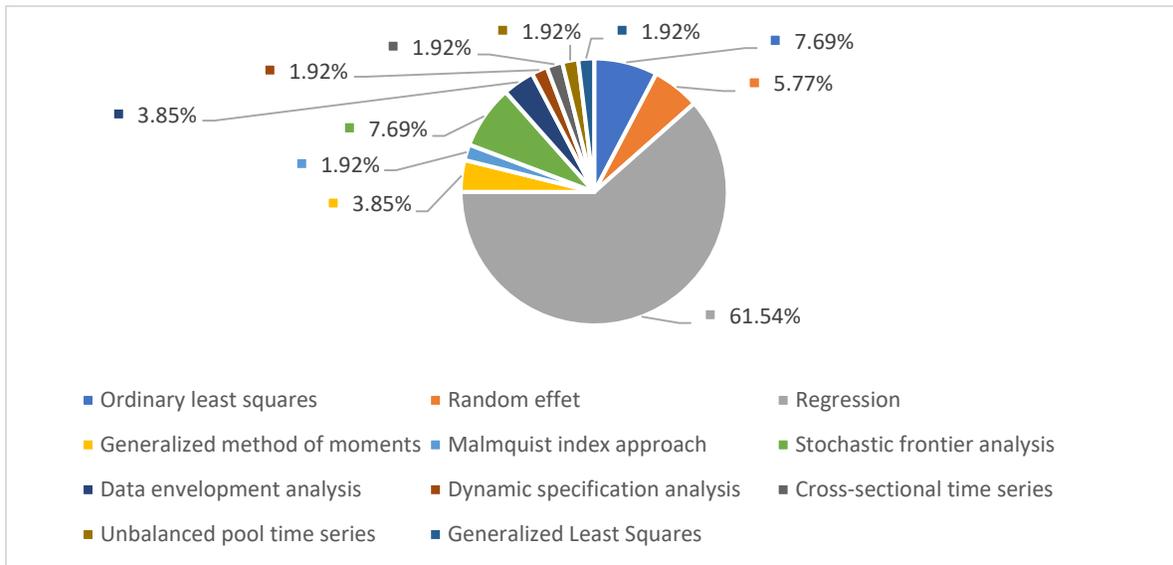

Source: Authors.

*Figure 4.* Econometric model usage by the sampled literature

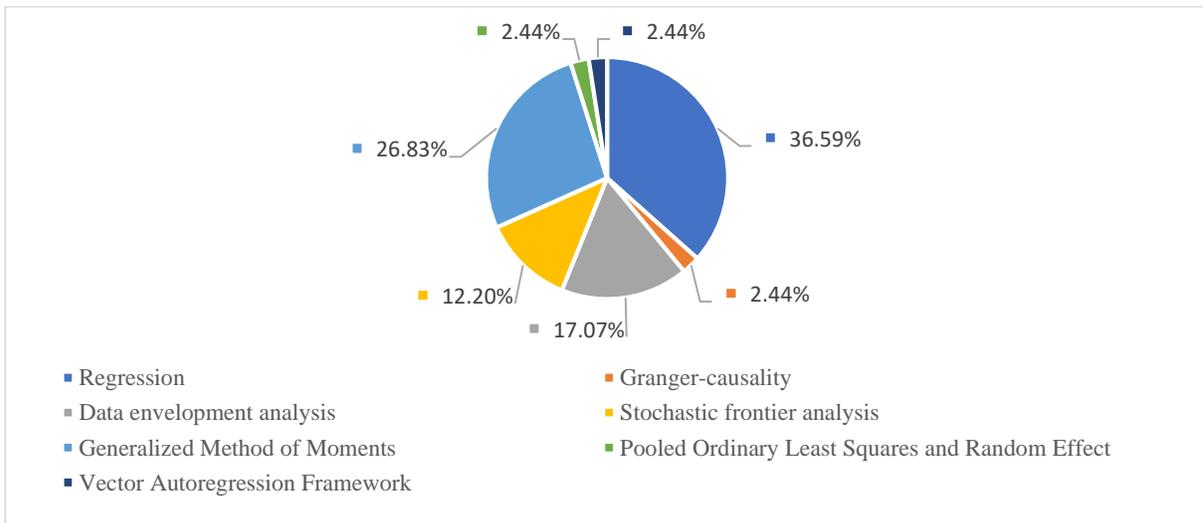

Source: Authors. Data from Alam et al. 2021.

*Figure 5.* Econometric model usage of top 20 most cited studies

We can conclude that regression models are the most used in the field of bank performance.

**Empirical methods**

According to Alam et al. (2021), the top 20 most cited studies used quantitative methods. This is also true for the sampled literature. Researchers seek to produce generalizable results, which is why data and econometric models are paramount.

**Literature review**

This section summarizes our analysis of the sampled literature. We have segmented it by type of determinant: exogenous and endogenous.

*Bank's exogenous determinants of performance*

This subsection shows the bank's exogenous determinants of performance.

Market concentration

Many authors affirm a positive correlation between bank performance and market concentration (Chamberlin & Wells, 1933; Mason, 1939; Bain, 1968; Bourke, 1989; Jeon & Miller, 2005; Athanasoglou et al., 2006; Menicucci & Paolucci, 2016; Almoneef & Samontaray, 2019). The market dictates the behaviour of banks, which then predicts performance. Others claim the opposite (Emery, 1971; Demsetz, 1973; Ross, 1977; Baumol et al., 1982; Berger, 1995; Anarfi et al., 2016; Otero et al., 2020). Instead, a bank must enhance its performance by reducing operating costs and maximizing profits. However, banks with few competitors behave competitively if the barriers to entry are weak (Baumol et al., 1982).

Inflation

Inflation can positively affect bank performance due to profits increasing more than costs. Banks are unaffected by the fluctuations of real GDP (Molyneux & Thornton, 1992; Athanasoglou et al., 2006; Jara-Bertin et al., 2014; Derbali, 2021). They benefit from their customers' failure to predict future inflation rates and act upon that information to boost their performance.

Other authors justify the negative impact of inflation by the inadequate responses of central banks. They resort to erratic policies to attempt to control inflation (Friedman, 1977). Higher inflation rates penalize agents who opt for saving. Consequently, there would be a higher incentive to benefit from the opportunity cost of investing in non-monetary assets. Inflation also harms the value of a currency. Exchange rates fluctuate unfavourably for banks, which will negatively impact their performance (Kosmidou, 2008; Umar et al., 2014). Chouikh and Blagui (2017) found that inflation has a negative impact. Alper and Anbar (2011) have revealed that inflation has no impact.

Interest rate

Interest rate directly affects the Net Interest Margin (NIM). A higher interest rate predicts a higher ROA and ROE, both of which predict bank performance (Molyneux & Thornton, 1992; Alper & Anbar, 2011; Nessibi, 2016; Derbali, 2021). However, Ogunbiyi and Ihejirika (2014) haven't found any correlation between bank performance and real interest rate.

GDP growth rate

GDP growth impacts bank performance positively (Sufian & Habibullah, 2009; Jara-Bertin et al., 2014; Caporale et al., 2015; Chen & Lu, 2020; Jreisat & Bawazir, 2021): economic agents performing better during periods of economic expansion. Pasiouras and Kosmidou (2007) and Chouikh and Blagui (2017) suggest the opposite. Alternatively, Derbali (2021) and Isnurhadi et al. (2021) claim that GDP growth has no impact. Moreover, the correlation between bank performance and economic growth behaves differently in developing countries (Ceylan & Ceylan, 2020).

Regulations

Regulations allow authorities to prevent market failures by rooting out fraud. It affects bank performance positively (Lozano-Vivas & Pasiouras, 2010; Yang et al., 2019). However, Yang et al. (2019) have pointed out that this impact is lower within emerging countries. While more strict regulations allow the banks to operate under less financial distress and a more secure framework, this limits the banks' activity and push their investment towards lower return assets (Pasiouras et al., 2009).

*Endogenous determinants of bank performance*

This section shows the bank's endogenous determinants of performance.

Bank size

With better access to the market and the capacity to generate economies of scale, a large bank tends to outperform a smaller one (Bonin et al., 2005; Athanasoglou et al., 2006; Jara-Bertin et al., 2014; Menicucci & Paolucci, 2016; Bahyaoui, 2017; Chouikh & Blagui, 2017; Elouali & Oubdi, 2018; Almoneef & Samontaray, 2019; Akoi & Andrea, 2020; Jreisat & Bawazir, 2021). However, Berger and Mester (1997) and Anarfi et al. (2016) claim that size has no bearing on performance. Some authors suggest a negative impact between size and performance due to higher management costs (Seiford & Zhu, 1999; Bahyaoui, 2017; Derbali, 2021; Al-Matari, 2023).

Ownership concentration

Certain authors think ownership concentration has a negative impact on bank performance (Jensen & Meckling, 1976; Goddard et al., 2004; Jaafar & El-Shawa, 2009; García-Meca & Sánchez-Ballesta, 2011; Al-Amarneh, 2014; Jarbou et al., 2018). This is because the top management can pursue strategies of undue profit appropriation. This causes a misallocation of profits and a decline shareholder value (Gedajlovic & Shapiro, 1998).

State-owned banks perform poorly versus private banks (Micco et al., 2007; Berger et al., 2009; Chouikh & Blagui, 2017). Non-listed banks outperform public banks (Bahyaoui, 2017). Moreover, banks with foreign shareholders tend to perform better than domestically-owned banks (Lee & Kim, 2013).

Capital structure

Higher equity means lower debt and more financial stability (Jara-Bertin et al., 2014; Islam & Nishiyama, 2016; Saleh & Abu Afifa, 2020; Isnurhadi et al., 2021). The bankruptcy cost theory by Berger (1995) states that banks anticipate the costs that would follow a potential bankruptcy. They opt for more equity to enhance their performance. However, Berger and Bouwman (2013) have found a negative correlation between equity and bank performance.

Cost efficiency

Cost efficiency is an indicator of operational processes mastery (Pasiouras & Kosmidou, 2007; Bahyaoui, 2017; Otero et al., 2020; Isnurhadi et al., 2021). A company is competitive and performant when it maximizes profits while minimizing costs. Beck et al. (2013) have determined that Islamic and conventional banks do not differ fundamentally in business or financial indicators. However, Chouikh and Blagui (2017) have determined that cost efficiency has no impact on bank performance.

Revenue diversification

There is no unanimity regarding its impact on bank performance. Revenue diversification allows the bank to diversify revenue streams and reduce risks (Alper & Anbar, 2011; Jara-Bertin et al., 2014; Jashim Uddin et al., 2022). However, Dietrich and Wanzenried (2014) have found that the optimal choice for banks is to rely on interest-based income because it outclasses other types. Revenue diversification becomes viable when interest margins decrease and financial markets increase.

Liquidity

Liquidity is the capacity to pay liabilities in the short- or the long-run. It is a positive bank performance determinant (Bourke, 1989; Jara-Bertin et al., 2014; Islam & Nishiyama, 2016; Elouali & Oubdi, 2018; Saleh & Abu Afifa, 2020). However, excessive liquidity can constitute a latent loss for the bank due to the opportunity cost incurred of holding liquidity instead of investing (Molyneux & Thornton, 1992; Goddard et al., 2004; Bahyaoui, 2017).

Credit risk

The focus on the issue of credit risk management increased substantially after the subprime crisis on 2008-2009 (Landskroner & Paroush, 2008). Almekhlafi et al. (2016) have determined that NPLs have a negative effect on bank performance. They are sunk costs, and endeavours to recover them would result in bigger losses. Prudent credit risk management positively impacts bank performance because it limits the probability of having doubtful receivables and defaulting clients (Kolapo et al., 2012; Almekhlafi et al., 2016).

Syed (2017) claims that there is a negative correlation between credit risk management and bank. His findings suggest that NPLs have an insignificant correlation with bank performance.

Credit risk management measures and processes are inversely associated with both the return on assets and the ROE.

**Variable usage and proxies**

This section summarizes the determinants used throughout the literature and the proxies used to measure them in table 2.

Table 2.

*Determinants of bank performance, proxies and sample of the literature*

| Determinants | Proxy | Sample of the literature |
|---|---|---|
| *Exogenous determinants* | | |
| Market concentration | Herfindahl-Hirschman Index: $$\sum_{i=1}^{n} Market\ Share_i^2$$ $$\frac{\sum Assets\ of\ the\ 3\ to\ 5\ largest\ banks}{Total\ assets\ of\ all\ banks\ in\ the\ country}$$ | Berger & Mester, 1997; Bourke, 1989; Elouali & Oubdi, 2018; Isnurhadi et al., 2021; Jashim Uddin et al., 2022 |
| Inflation rate | Average annual inflation rate. | Molyneux & Thornton, 1992; Athanasoglou et al., 2006; Kosmidou, 2008; Alper & Anbar, 2011; Umar et al., 2014; Chouikh & Blagui, 2017; Derbali, 2021 |
| Interest rate | Lending rate used by the bank. | Molyneux & Thornton, 1992; Alper & Anbar, 2011; Ogunbiyi & Ihejirika, 2014; Nessibi, 2016; Derbali, 2021 |
| GDP growth rate | Annual growth rate of GDP. | Pasiouras & Kosmidou, 2007; Sufian & Habibullah, 2009; Caporale et al., 2015; Chouikh & Blagui, 2017; Chen & Lu, 2021; Derbali, 2021; Isnurhadi et al., 2021; Jreisat & Bawazir, 2021 |
| Regulations | The three pillars of Basel III: 1. Capital Adequacy Ratio: $$\frac{Tier\ 1 + Tier\ 2 + Tier\ 3 \pm Adjustments}{Risk\ Weighted\ Assets}$$ 2. ICAAP certification (Internal Capital Adequacy Assessment Process); 3. Public disclosure & transparency. | Lozano-Vivas & Pasiouras, 2010; Yang et al., 2019 |
| *Endogenous determinants* | | |
| Bank size | $\ln (Total\ Assets)$ $\ln (Accounting\ Value\ of\ the\ Bank)$ $\dfrac{Bank\ Turnover_i}{\sum_{i=1}^{n} Bank\ Turnover_i}$ $\ln^2 (Total\ Assets)$ | Alper & Anbar, 2011; Berger & Bouwman, 2013; Islam & Nishiyama, 2016; Jarbou et al., 2018; Almoneef & Samontaray, 2019; Saleh & Abu Afifa, 2020; Akoi & Andrea, 2020; Anarfi et al., 2016; Bahyaoui, 2017; Jreisat & Bawazir, 2021 |
| Ownership concentration | Percentage of shares owned by the largest shareholder. Percentage of shares owned (minimum 5%) by the N largest shareholder (maximum 5). Percentage of state-owned, private-owned and foreign equity. | Jarbou et al., 2018; Jaafar & El-Shawa, 2009; Micco et al., 2007 |
| Capital structure | Equity to total assets: $$\frac{Equity}{Total\ Assets}$$ Capital adequacy ratio: $$\frac{Tier\ 1 + Tier\ 2}{Risk\ Weighted\ Assets}$$ | Berger & Bouwman, 2013; Dietrich & Wanzenried, 2014; Islam & Nishiyama, 2016; Bahyaoui, 2017; Saleh & Abu Afifa, 2020; Jreisat & Bawazir, 2021; Isnurhadi et al., 2021; Jashim Uddin et al., 2022; Berger, 1995 |
| Cost efficiency | Cost-to-income ratio: $$\frac{Costs}{Income}$$ Operating expenses to staff ratio: $$\frac{Operating\ Expenses}{Employees}$$ Operational efficiency ratio: $$\frac{Operating\ Expenses}{Net\ Income}$$ Optimal cost ratio: $$\frac{Optimal\ Total\ Cost}{Total\ Cost}$$ $$\frac{Total\ Overhead\ Costs}{Total\ Assets}$$ | Dietrich & Wanzenried, 2014; Saleh & Abu Afifa, 2020; Isnurhadi et al., 2021;Islam & Nishiyama, 2016; Bahyaoui, 2017; Otero et al., 2020; Jashim Uddin et al., 2022 |
| Revenue diversification | Interest income share: $$\frac{Non\ Interest\ Income}{Total\ Income}$$ Non-interest income to assets: $$\frac{Non\ Interest\ Income}{Total\ Assets}$$ Non-interest income vs. interest income: $$\frac{Non\ Interest\ Income}{Interest\ Income}$$ Non-interest income to assets: $$\frac{Non\ Interest\ Bearing\ Assets}{Total\ Assets}$$ | Dietrich & Wanzenried, 2014; Jashim Uddin et al., 2022; Alper & Anbar, 2011; Jreisat & Bawazir, 2021; Jashim Uddin et al., 2022 |
| Liquidity | $$\frac{Total\ Receivables}{Balance\ Sheet\ Total}$$ $$\frac{Liquid\ Assets}{Deposits + Short\ Term\ Funding}$$ $$\frac{Total\ Loans}{Total\ Assets}$$ | Bahyaoui, 2017; Islam & Nishiyama, 2016; Jreisat & Bawazir, 2021 |
| Credit risk | $$\frac{Loan\ Loss\ Provisions}{Total\ Loans}$$ $$\frac{Doubtful\ Loans}{Total\ Loans}$$ | Dietrich & Wanzenried, 2014; Saleh & Abu Afifa, 2020; Bahyaoui, 2017 Berger & Bouwman, 2013;Alper & Anbar, 2011; Jreisat & Bawazir, 2021 |

$$\frac{Risk\ Weighted\ Assets}{Gross\ Total\ Assets}$$

$$\frac{Loans\ under\ followup}{Total\ Loans}$$

$$\frac{Impaired\ Loans}{Gross\ Loans}$$

Source: Authors

**Future research avenues**
In this section, we will discuss why further research into the determinants of bank performance is relevant.

*Deepen research on technology-driven determinants of bank performance*
The literature review shows an abundance of studies and research on the exogenous and endogenous determinants of bank performance. However, we have noticed few studies have tackled the topic of technology-driven determinants of bank performance.

Growth of FinTech start-ups
COVID-19 catalysed the development of FinTechs. They became increasingly successful and powerful, to the point of competing with banks. This resulted in two perspectives: FinTechs could be an opportunity or a threat in a bank's strategy. The latter acquires the former to eliminate a potential competitor and gain more expertise.

There have been few research endeavours on the impact of the growth rate of FinTechs on bank performance (Phan et al., 2020; Zhou et al., 2021). Therefore, the growth of FinTechs' impact on bank performance deserves further research.

Investment in digitalization
Traditional banking relied on clients going to branches to benefit from banking services. However, COVID-19 has significantly accelerated the development of digital tools and processes, especially in the banking industry. Digitalization has become an increasingly decisive factor since then. Yet, the literature does not cover this aspect of bank performance sufficiently. There are few empirical studies on digitalization (Forcadell et al., 2020; Abou-foul et al., 2021; Coryanata et al., 2023). Therefore, the impact of banks' investment in digitalization deserves further research.

*Extend research to other regions*
Most sampled studies were conducted in Europe or on a global scale. It is the opposite for emerging regions, namely MENA which has 2 occurrences. The articles that focused on MENA studied specific aspects of the determinants of bank performance. The same can be observed for the 20 most cited articles.

This gap should be addressed to provide new insight to researchers and policymakers, considering how the MENA region is a core player in the global economy.

*Provide insights using data before and after COVID-19*

The literature is aware of the changes that COVID-19 has caused in the banking industry. However, there is a scarcity of articles that have studied the determinants of bank performance after COVID-19. We aim to provide research by retaining more data range than the studies we have reviewed.

*Introducing a dual-approach to the empirical methodology*

The relevant literature uses quantitative methods to produce results. This allows researchers to obtain generalizable conclusions. We aim to utilize both quantitative and qualitative methods for broader perspectives.

## Conclusion

Bank performance is a crucial subject for researchers, academicians and professionals from the banking industry. They have endeavoured to study the factors affecting the former.

The literature review has allowed us to identify gaps within: extending the research to emerging regions or counties that haven't been sufficiently studied. This paper has also revealed a crucial future research avenue: technology-driven factors of bank performance by the COVID-19 pandemic in the banking sector.

In light of the relevance and added value of this paper, we believe it will be an excellent addition to the finance and banking literature. Analysing the contributions of past studies has paved the way for future research avenues.

## Acknowledgments

Not applicable.

## Funding

## Conflict of Interests

No, there are no conflicting interests.

## Open Access